\documentclass[epj,twocolumns]{svjour}
\usepackage{graphics}
\begin{document}
\title{Polarization catastrophe in the polaronic Wigner crystal}

\author{S. Fratini \and P. Qu\'{e}merais}
%
\offprints{quemerai@lepes.polycnrs-gre.fr}
%
\institute{Laboratoire d'Etudes des Propri\'{e}t\'{e}s Electroniques des
Solides\\
CNRS, BP 166, 38042 Grenoble Cedex 9, France}
\date{Received: date / Revised version: date}
%
\abstract{
We consider a three dimensional Wigner crystal of electrons 
lying in a host ionic dielectric. Owing to their interaction with the lattice 
polarization, each localized electron forms a polaron.
We study the collective excitations of such a polaronic Wigner crystal at
zero temperature, taking into account the quantum fluctuations of the
polarization within the Feynman harmonic approximation. We show
that, contrary to the ordinary electron crystal,  the system undergoes a
polarization catastrophe when the density is increased.
An optical signature of this instability is derived, whose trend agrees 
with the experiments carried out in Nd-based cuprates. 
\PACS{
      {PACS-key}{describing text of that key}   \and
      {PACS-key}{describing text of that key}
     }
}
%
\authorrunning{S. Fratini and P. Qu\'emerais}
\titlerunning{Polarization catastrophe in the polaronic Wigner crystal}
\maketitle

\section{Introduction}
\label{sec:Introduction}

Let us consider a Wigner crystal (WC) of electrons --- a state where the
carriers form an ordered array in order to minimize the repulsive
Coulomb energy --- and put it
into a polar dielectric --- a material 
which can respond to the electron motion by displacing the positive and
negative ions.
If the density of charges added to the dielectric is low, 
each carrier independently forms a \textit{polaron}, i.e. it is
surrounded by a polarization cloud
due to the interaction with the longitudinal optical phonon modes of
the host medium. In the opposite high density limit, dynamical
screening will prevent both the localization of charges due to the
Coulomb repulsion and the formation of polaronic bound states, 
thus leading to a weakly interacting liquid. The phase transition
occurring upon varying the density between these two extreme limits
has been 
analysed in ref. \cite{PWC} starting from the crystallized state.
The resulting picture allows to
distinguish between two qualitatively different situations. 
While in the weak and intermediate electron-phonon coupling regimes, 
the polaron Wigner
crystal (PWC) can melt towards a polaron liquid, this is 
prevented when the polar coupling is strong: in that case the
polarization cloud is virtually frozen and polarons tend to 
\textit{dissociate} as the system becomes metallic. 

The scope of this work is to go beyond the
mean-field approach presented in ref.   \cite{PWC},
with particular attention to the collective modes arising from the
long-range interactions between electrons in the crystallized
phase.  The main result of our study is
that in the strong electron-phonon coupling regime, 
where polarons have a well-defined internal structure, 
the dipolar electron-electron interactions
can lead to  very anomalous dielectric properties, and eventually to 
an instability at high enough density, characterized by a
softening of a long wavelength transverse collective mode. An
equivalent result was already obtained in previous works 
assuming a static polarization potential \cite{cata}, i.e. neglecting
the quantum fluctuations of the polarization. Similar indications have
also been given by other authors \cite{lorenzana} starting  from a
classical liquid state. 
All these cases are
practical realizations of the
very general phenomenon of a ``dielectric catastrophe'' \cite{textbook}, 
occurring when local dipoles oscillating with a given
restoring force (the electrons inside the polarization
potential-wells) interact through the long-range Coulomb
forces, which tend to soften the transverse 
restoring potential. Such instability, which can be traced back to the
local Lorentz field,
has an experimental signature: the
frequency of the polaron peak in the optical conductivity
decreases  with doping, a trend that has indeed been observed
in underdoped  Nd$_{2-x}$Ce$_x$CuO$_{4-y}$ (NCCO) below the
insulator-to-metal transition \cite{calvani}.

The paper is organized as follows. 
The main features of the model, which was thoroughly introduced in
reference  \cite{PWC}, are summarized in
section \ref{sec:modele}. In section \ref{phonon}, we calculate the
phonon dispersion of the PWC on the basis of Feynman's harmonic trial model
for the polarons. 
In section \ref{dielectric constant}, we evaluate the effect of the 
vibrations of the PWC by
calcultating the dielectric constant $\varepsilon \left( {\bf k}, \omega
\right)$ as a function of the density. We show that it
is \textit{negative} on a large region of $\mathbf k$ and $\omega$ 
when the system approaches the instability.
In conclusion,
we propose a scenario where free carriers 
liberated by the dissociation of polarons could
coexist with the residual localized polarons in a ``mixed'' phase 
beyond the instability, and speculate about a novel pairing mechanism 
where the collective
vibrations of the PWC or, equivalently, the ``overscreening'' related to the
negativity of the dielectric constant $\varepsilon(k,\omega)$, acts as
a glue for the
Cooper pairs.



\section{Model and approximations}
\label{sec:modele}

The model for the crystallized state has been thoroughly defined 
in ref. \cite{PWC}, equations (1)-(5). Electrons with band mass
$m^*$ at a density $n=\left(4\pi /3\right)^{-1}R_s^{-3}$ interact 
directly through the long-range Coulomb forces. These are immersed in 
a polar background characterized by the longitudinal  
phonon frequency $\omega_{LO}$ and the effective dielectric constant
$\tilde{\varepsilon}$. The polar medium  responds to the electron motion
through a Fr\"ohlich type interaction characterized by a coupling
constant  $\alpha = \left( m^*/ 2 \hbar^3
  \omega_{LO}\right)^{1/2} e^2 / \tilde{\varepsilon}$, which can lead to
polaron formation. 
Once the degrees of
freedom of the polarization field are integrated out, the resulting action for
the electrons can be expanded assuming small displacements ${\bf u}_i$
relative to the equilibrium positions. To 
quadratic order, this reads:
\begin{equation}
S\left( \left\{ {\bf u}_{i}\right\} \right) =\sum_{i}S_{i}+\frac{1}{2}\sum_{i
\neq j}S_{ij}  \label{Sfinal}
\end{equation}
with
\begin{eqnarray}
S_{i} &=&-\beta \frac{9e^{2}}{10\varepsilon _{s}R_{s}}-\int_{0}^{\beta
} \left[ \frac{m^{*}}{2}\dot{u}_{i}^{2}\left( \tau \right) +\frac{m^*}{2}
\frac{\omega _{W}^{2}}{\varepsilon _{s}}u_{i}^{2}\left( \tau \right)
\right] d\tau  \nonumber \\
&&+\frac{\omega _{LO}e^{2}}{4{\tilde{\varepsilon}}}{\int_{0}^{\beta } \!
{\int_{0}^{\beta }{\frac{G_{\omega _{LO}}(\beta ,\tau -\sigma )}{|{\bf u}
_{i}(\tau )-{\bf u}_{i}(\sigma )|}d\tau d\sigma }}}  \label{S-diag}
\end{eqnarray}
and
\begin{eqnarray}
& &{S_{ij} =-\frac{e^{2}}{{\varepsilon }_{\infty }}
\sum_{\alpha \gamma}\int_{0}^{\beta }
u_{i}^{\alpha }( \tau) \Lambda _{ij}^{\alpha \gamma }
u_{j}^{\gamma }( \tau)  d\tau} \label{S-off-diag} \\
& &+\frac{\omega _{LO}e^{2}}{2{\tilde{\varepsilon}}}
\sum_{\alpha \gamma}\int_{0}^{\beta } \! \int_{0}^{\beta }
u_{i}^{\alpha }(\tau)\Lambda _{ij}^{\alpha \gamma }
u_{j}^{\gamma }(\sigma)
G_{\omega_{LO}}(\beta ,\tau -\sigma )
d\tau d\sigma   \nonumber
\end{eqnarray}
where the indices $\alpha ,\gamma =(x,y,z)$ denote the cartesian
coordinates. The dipolar matrix elements are defined as:
\begin{equation}
\Lambda _{ij}^{\alpha \gamma }=\frac{\delta _{\alpha \gamma
}R_{ij}^{2}-3R_{ij}^{\alpha }R_{ij}^{\gamma }}{R_{ij}^{5}}.
\end{equation}
We have introduced the phonon propagator
$\mbox{$G_{\omega }(\beta ,\tau -\sigma )$}=(\bar{n}
+1)e^{-\omega |\tau -\sigma |}+\bar{n}e^{\omega |\tau -\sigma |}$,  together
with $\bar{n}=(e^{\beta \omega }-1)^{-1}$.
The frequency  $\omega_W$ is related to
the  plasma frequency
by the relation $\omega_W^2=\omega_p^2/3=e^2/m^*R_s^3$. $\beta$ is
the inverse temperature, and the units are such that
$\hbar = k_B = 1$.

Expressions (\ref{Sfinal})-(\ref{S-off-diag}) constitute the basic
model for the polaron Wigner crystal (PWC). It 
is based essentially
on  the following approximations: (i) anharmonic terms
in the expansion in $\{{\bf u}_i\}$ are not considered; (ii) 
exchange effects between
electrons are neglected, and so is
the possible (weak) magnetic ordering of the crystal.
For the main purposes of our studies, i.e. the crystal stability, all
these approximations are well justified in the range of densities that we
consider, as was discussed in detail in ref.\cite{PWC}.

The local part of the electron action has been extensively
studied in ref. \cite{PWC}, by isolating neutral entities containing
only one electron --- the so-called
Wigner spheres --- which are non-interacting at mean-field level. In
this approach, all
the correlation effects are carried by a local harmonic potential [the
second term in square brackets in eq. (\ref{S-diag})], which merely
adds to the polaron problem
(the additional constant term is the electrostatic energy coming from the
jellium approximation).  
The last term in eq. (\ref{S-diag}) is the
self-induced interaction mediated by  
the polarization, which is responsible for polaron formation.
It is not solvable, but it can be treated successfully
within the Feynman
variational approach \cite{feynman} by introducing a quadratic trial
action $S_0$ of the form
\begin{eqnarray}
S_{0} &=&-\frac{m^{*}}{2}\int \dot{ u}^{2} d\tau -\frac{m^{*}}{2}\frac{
\omega _{W}^{2}}{\varepsilon _{s}}\int  u^{2} d\tau  \nonumber \\
&-&\frac{Kw}{8}\int \int {|{\bf u}(t)-{\bf u}(s)|}
^{2}G_{w}(\beta ,\tau -\sigma )d\tau \,d\sigma ,  \label{qaction}
\end{eqnarray}
(we have dropped the site index, since at this stage
all the sites are equivalent). It can be seen that expression
(\ref{qaction}) results from the path-integration over
${\bf X}$ of the following Lagrangian:
\begin{equation}
L_{0}=\frac{m^{*}}{2}{\dot{u}}^{2}+\frac{M}{2}{\dot{X}}^{2}-\frac{K
}{2}\left( {\bf u-X}\right) ^{2}-\frac{m^{*}}{2}\frac{\omega _{W}^{2}}{
\varepsilon _{s}}{u}^{2}.  \label{L_0}
\end{equation}
where $w^2=K/M$.
In eq. (\ref{L_0}), $K$ and $w$ are variational parameters to be
adjusted in order to minimize the free energy.
This method of approximation,  not only yields the
best known analytical results for the polaron energy, but it also  
gives a very good
insight into the physics of the polaron problem:
when the variational parameters $K$ and $w$ are properly adjusted,
the model (\ref {L_0}) is a good schematization of the true polaron,
especially at low temperatures (roughly at $T< \omega_{LO}$, where
the polarization cloud is sufficiently ``rigid'').
Basically, the actual polarization of the polaron is replaced by an
auxiliary  particle with coordinate ${\bf X}$ and mass $M$, which reduces the
many-body polaron problem to a two-body problem.

Let us now consider the non-local part (\ref{S-off-diag}) for $i \neq
j$. These terms, which represent dipole-dipole interactions between
different electrons,  become important beyond the mean-field approximation
described above, and they are responsible for the dipersion of the vibrating
collective modes.
The first term in eq. (\ref{S-off-diag}) is instantaneous,
and it comes out from the direct Coulomb electron-electron repulsion.
The second term is a
retarded dipole-dipole interaction which, as the polaron term in
(\ref{S-diag}), is mediated by the polarization. Its physical meaning is that
an electron $i$, when it moves, feels the polarization field created by
electron $j$. It is important to emphasize that, unlike the polaron
term in (\ref{S-diag}), this retarded dipole-dipole interaction vanishes
with $\omega_{LO}$ (i.e. $\alpha \rightarrow \infty$):
when the polarization is completely frozen, only the instantaneous
dipole-dipole interaction remains at work. This particular limit, which is very
instructive, has been already studied
in ref. \cite{cata}. Here, we consider the general case
$\omega_{LO}\neq 0$, where the
retarded dipole-dipole interactions cannot be neglected.

\subsection{Effective quadratic model}

In the same spirit as the Feynman approach for the single
polaron, we introduce an effective Lagrangian:
\begin{eqnarray}
L_{eff} &=&\sum_i{\frac{m^{*}}{2} \dot{ u_i}^{2} +\frac{M}{2}{\dot{X_i}}^{2}-
\frac{K}{2} \left( {{\bf u}_i-{\bf X}_i}\right) ^{2}- \frac{m^{*}}{2}
\frac {\omega _{W}^{2}}{\varepsilon _{s} }u_i^{2} } \nonumber \\
&-& \frac{1}{2} \sum_{i \neq j} {\bf u}_i \hat{A}_{ij}
{\bf u}_j - \sum_{i \neq j} {\bf u}_i \hat{B}_{ij} {\bf X}_j.
\label{Leff}
\end{eqnarray}
where the $\hat{A}_{ij}$ and $\hat{B}_{ij}$
are $3\times 3$ matrices in coordinate space.
This model can be thought of as
a generalization of the Lorentz lattice of dipoles,
which includes retarded dipolar interactions through the auxiliary particles
${\bf X}_i$.

The effective action for the electrons is obtained after path-integration on
the coordinates ${\bf X}_i$:
\begin{eqnarray}
S_{eff} &=& - \frac {m^*}{2} \sum_i \int d\tau   [ \dot{u}_i^2(\tau)
-  (K / m^* + \omega_W^2 / \varepsilon_s  ) u_i^2(\tau) ] \nonumber \\
&-& \frac{1} {2} \sum_{i\neq j}
\int {\bf u}_i(\tau) \hat {A}_{ij} {\bf u}_j(\tau) d\tau \nonumber  \\
&+& {\frac {1}{4Mw}}  \int\! \! \int d\tau d\sigma  G_{w}(\beta,\tau-\sigma)
\times \label{Seff} \\
&\times& \left\lbrace \sum_i {\bf u}_i(\tau) [ K^2
-2K \hat{B}_{ii}+ \sum_{l} \hat{B}_{li} \hat{B}_{li} ]
 {\bf u}_i (\sigma) \right. \nonumber \\
& & \left.  \sum_{i \neq j} {\bf u}_i(\tau) \left[
-2K \hat{B}_{ij} + \sum_{l} \hat{B}_{li} \hat{B}_{lj} \right] {\bf u}_j
(\sigma) \right\rbrace \nonumber
\end{eqnarray}
\label{approximations}
At this stage, the most rigorous way to proceed would be to
self-consistently solve the full Feynman variational problem given by:
\begin{equation}
E= \min_{K,w} [ E_0 + \lim_{\beta \rightarrow \infty} <S-S_{eff}>],
\label{Emin}
\end{equation}
where $E_0$ is the zero-point energy of all the eigenmodes of
(\ref{Leff}) \cite{note-Salje}.
However, we can get more insight into the problem by
making use of two additional approximations:

\textit{(i)} We restrict to a single variational parameter, by putting
$w = \omega_{LO}$ in the expression (\ref{Seff}). As was pointed out
in ref. \cite{PWC}, this has very little effect on the results
when dealing with moderate to large e-ph couplings ($\alpha > \sim
5$). From now on, we will mainly focus on this case, which is to our
opinion the most interesting one.

\textit{(ii}) For the variational parameter $K$, we use
the results obtained in the framework of the mean-field
approximation (see ref. \cite{PWC}). The underlying idea is that of a
perturbation expansion around the mean-field result, so that the first
correction is calculated by taking the zeroth order as a good starting
point. At low enough density, the
dipolar terms are indeed a small perturbation, since they explicitely
involve  the fluctuations ${\bf u}_i$ of the electrons around their
equilibrium positions, which are small in this limit. More
quantitatively, in the case of the ordinary WC, it was shown that
the inclusion of  dipolar interactions changes
the total energy by no more than 5-10 percent \cite{carr}.

Of course, the previous
argument no longer applies when approaching the critical
density, where one can expect some quantitative deviations from the
predicted phonon spectrum. However, our results should be at least
qualitatively correct in all the insulating phase.

Within this framework, there is no need to perform new path-integrations:
the matrices $\hat{A}_{ij}$ and
$\hat{B}_{ij}$ are chosen in order that all the dipole-dipole terms of
$S-S_{eff}$ in equation (\ref{Emin}), vanish term by term. This leads to
the following set of equations:
\begin{eqnarray}
\hat{A}_{ij} &=& \frac {e^2} {\varepsilon_{\infty}} \hat{\Lambda}_{ij}
\quad ; \quad i \neq j  \label{ABa}\\
-2K \hat{B}_{ii} + \sum_l \hat{B}_{li} \hat{B}_{li}& =& 0 \label{ABb} \\
-2K \hat{B}_{ij} + \sum_l \hat{B}_{li} \hat{B}_{lj} &=& K{\frac
{e^2}{\tilde \varepsilon}} \hat{\Lambda}_{ij} \quad ; \quad i \neq j.
\label{ABc}
\end{eqnarray}
The first equation straightforwardly defines the matrix $\hat{A}_{ij}$ of the
instantaneous dipole-dipole interactions,  and
the set of equations for the retarded terms $\hat{B}_{ij}$  will be solved
in the next section. As a result, we are able
to calculate the
phonon spectrum of the effective Lagrangian (\ref{Leff}), which in the
same manner as
(\ref{L_0}) for the single polaron problem, mimics the collective 
excitations of the PWC.

\section{Phonon Spectrum and Instability}
\label{phonon}

In this section, we derive the phonon spectrum of the Lagrangian
(\ref{Leff}), as was done for the
usual Wigner Crystal and Lorentz lattice of dipoles in references
 \cite{pines,bagchi,carr}.

First, the WC of electrons is recovered from eq. (\ref{Leff}) by putting
the e-ph coupling to zero, with $\varepsilon_s \rightarrow
\varepsilon_{\infty}$. All the terms involving $\{{\bf X}_i\}$
 thus vanish, and we obtain  \cite{bagchi}:
\begin{equation}
\label{L_W}
L_{W} =\sum_i \frac{m^{*}}{2} \dot{ u_i}^{2}-\frac{m^{*}}{2}
\frac{\omega _{W}^{2}}{\varepsilon _{\infty}}u_i^{2} -
\frac{1}{2} \sum_{i \neq j,\mu \nu} {\bf u}_i \hat{A}_{ij}
{\bf u}_j
\end{equation}
with the matrices $\hat{A}_{ij}$ for $i \neq j$ given by
equation (\ref{ABa}). The potential term of (\ref{L_W}) can be
included in the definition of
the $\hat{A}$'s by putting $\hat{A}_{ii}=(m^* \omega_W^2/\varepsilon_{\infty})
\hat{I}$, with  $\hat{I}$ the identity matrix.
We can now diagonalize the coupled equations of electron
motion
\begin{equation}
m^* {\bf \ddot u}_i(t) - \sum_j \hat{A}_{ij} {\bf u}_j(t)=0
\end{equation}
by introducing eigenmodes of the form:
\begin{equation}
{\bf u}_i(t)= u_{{\bf k}\lambda} { \bf \epsilon}_{{\bf k} \lambda}
e^{i({\bf k}\cdot{\bf R}_{ij}-\frac{\omega({\bf k},
\lambda)}{\sqrt{\varepsilon_\infty}} t)}.
\end{equation}
where ${ \bf \epsilon}_{{\bf k} \lambda}$ is the polarization vector of the
mode $({\bf k},\lambda)$, ${\bf k}$ being a wavevector in the Brillouin
zone, and $\lambda$ the index of the (three) phonon branches.
If we define the Fourier transform as
\begin{equation}
\hat{A}_{\bf k}= \frac {1} {N} \sum_j \hat{A}_{ij} e^{i{\bf k}\cdot{\bf
R}_{ij}}.
\label{four}
\end{equation}
then the eigenfrequencies fulfill:
\begin{equation}
{\bf \epsilon}_{{\bf k} \lambda}
N \hat{A}_{\bf k} {\bf \epsilon}_{{\bf k} \lambda^\prime}=
m^*\frac{\omega^2({\bf k},
\lambda)}{\varepsilon_\infty} \delta_{  \lambda \lambda^\prime}
\label{eigen}
\end{equation}
For a given Bravais lattice,
the matrix elements of $\hat{A}_{\bf k}$ are given by \cite{pines}:
\begin{eqnarray}
A_{\bf k}^{\mu \nu} =k^{\mu} k^{\nu} V_{\bf k}+\sum_{{\bf K} \neq 0}
V_{\bf k+K}(k^{\nu}+K^{\nu})(k^{\mu}+K^{\mu}) \nonumber \\
- \sum_{{\bf K} \neq 0} K^{\mu} K^{\nu} V_{\bf K}.
\label{ak}
\end{eqnarray}
where $V_k=4 \pi e^2 /k^2 \varepsilon_{\infty}$
is the Fourier transform of the Coulomb potential, with $V_{k=0}$ set to zero
for charge neutrality, and ${\bf K}$ a
reciprocal lattice vector. From (\ref{eigen}) and
(\ref{ak}) it is easy to prove the following result:
\begin{equation}
\sum_{\lambda} \frac{\omega({\bf k},\lambda)^2}{\varepsilon_\infty}
=3 \omega_W^2 / \varepsilon_{\infty}
= \omega_p^2 /
\varepsilon_{\infty}\label{kohn}
\end{equation}
which is known as the Kohn sum rule \cite{pines,bagchi}. The phonon
spectrum of the WC
contains two acoustical transverse modes ($t$) and one optical longitudinal
mode ($\ell$), which satisfies $\omega(0, \ell)=\omega_p /
\sqrt{\varepsilon_{\infty}}$.

Let us now calculate the Fourier transform $\hat{B}_{\bf k}$ of the matrices
$\hat{B}_{ij}$, defined in the same way as $\hat{A}_{\bf k}$ in eq.
(\ref{four}). Multiplying
eq. (\ref{ABc}) by $\exp (i{\bf k} \cdot {\bf R}_{ij})$, summing over
$j$, then making use of (\ref{ABb}),
one gets:
\begin{equation}
-2KN\hat{B}_{\bf k}+N\hat{B}_{\bf k}N\hat{B}_{\bf k}= K \frac {e^2}
{\tilde \varepsilon} N \hat{\Lambda}_{\bf k} - \frac{m^* \omega_W^2}{\tilde
\varepsilon} \hat{I}
\label{bk}
\end{equation}
The above equation can be solved in the same
basis $\{{\bf \epsilon}_{{\bf k},\lambda}\}$ which diagonalizes
$\hat{\Lambda}_{\bf k}$ and  $\hat{A}_{\bf k}$. Therefore, one obtains
${\bf \epsilon}_{{\bf k},\lambda} N \hat{B}_{\bf k} {\bf \epsilon}_{{\bf
k}, \lambda \prime}= g({\bf k}, \lambda) \delta_{\lambda,
\lambda \prime}$ where:
\begin{equation}
g({\bf k}, \lambda)= K \left[ 1 - \sqrt{ 1 + {\frac {1} {\tilde
\varepsilon}} {\frac {m^*} {M}} {\frac {\omega^2 ( {\bf k}, \lambda)-
\omega_W^2} {\omega_{LO}^2}}} \right]
\label{gk}
\end{equation}

\subsection{Phonon spectrum}

We can now determine the whole phonon spectrum of the PWC
by diagonalizing the Lagrangian (\ref{Leff}). Let us introduce
the Fourier transforms
\begin{eqnarray}
{\bf u}_i &=& {\frac {1} {\sqrt {N}}} \sum_{{\bf k} \lambda}
 u_{{\bf k} \lambda} {\bf \epsilon}_{{\bf k} \lambda} e^{-i{\bf k}
 \cdot {\bf R}_i} \nonumber \\
{\bf X}_i &=& {\frac {1} {\sqrt {N}}} \sum_{{\bf k} \lambda}
 X_{{\bf k} \lambda} {\bf \epsilon}_{{\bf k} \lambda} e^{-i{\bf k}
 \cdot {\bf R}_i}.
\label{FT}
\end{eqnarray}
then the equations of motion  become
\begin{eqnarray}
\ddot{u}_{{\bf k} \lambda} &=& \left[ {\frac {\omega_W^2}
{\tilde \varepsilon}} - {\frac {\omega^2({{\bf k} \lambda})}
{\varepsilon_{\infty}}}  -\frac{K}{m^*} \right]
u_{{\bf k} \lambda} + \frac{\left[ K- g({{\bf
k}, \lambda}) \right]}{m^*} X_{{\bf k} \lambda} \nonumber \\
\ddot{X}_{{\bf k} \lambda} &=& \frac{\left[K-g({{\bf k} \lambda}) \right]}{M}
u_{{\bf k} \lambda} -\frac{K}{M} X_{{\bf k} \lambda}.
\label{eqmot}
\end{eqnarray}
For each wavevector $\bf{k}$ and polarization
$\lambda$, the solution of the secular equation gives the
eigenfrequencies $\Omega({\bf k}, \lambda)$ of the PWC in terms of the
eigenfrequencies $\omega({\bf k}, \lambda)$ (formula (18)) of the WC of
electrons:
\begin{eqnarray}
\Omega^2({\bf k}, \lambda)&=& \frac {1} {2}
 \left\lbrace \omega_{pol}^2 + \frac {
\omega^2({{\bf k}, \lambda})} {\varepsilon_{\infty}} \right.
\label{Eigen}\\
&& \left.
\pm \sqrt{ \left[\omega_{pol}^2+
\frac{\omega^2({{\bf k},\lambda})}{\varepsilon_\infty} \right]^2
-4 \omega_{LO}^2 \frac{\omega^2({{\bf k}, \lambda})}{ \varepsilon_s}}
 \right\rbrace \nonumber
\end{eqnarray}
where we have defined
\begin{equation}
\omega_{pol}^2=\left(
\frac{M}{m^*}+1\right)\omega_{LO}^2-\frac{\omega_W^2}{\tilde{\varepsilon}}
\label{wpol}
\end{equation}

In the following, we shall denote by $\Omega_{int}({\bf k}, \lambda)$
[$\Omega_{ext}({\bf k}, \lambda)$]
the solution  with the $+$ [$-$] sign. By looking at the equations of
motion (\ref{eqmot}), one can verify that in the limit of low $k$,
these frequencies are associated respectively with the out-of-phase
and in-phase vibrations of the electrons relative to the polarization.
Interestingly enough, within our approximation scheme,
the polaron effect now enters in the frequency spectrum through the
parameter $\omega_{pol}$ alone. Moreover, the parameter $\omega_{pol}$
is identified as the frequency of
the transverse mode $\Omega_{int}({\bf k}, t)$ at $k=0$.

\begin{figure}
\centerline{\resizebox{9cm}{!}{\includegraphics{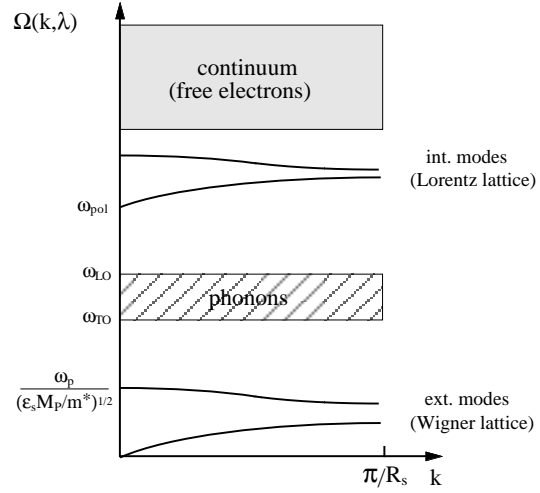}}}
\caption{The excitation spectrum of the PWC.}
\label{fig-Eigen}
\end{figure}
It is instructive to work out the solutions of eq. (\ref{Eigen}) in
the low density limit, where $\omega_W$ is small compared to the
internal frequency of an isolated polaron. In this case we obtain
\begin{eqnarray}
\Omega^2_{ext}({\bf k}, \lambda) &\simeq& \frac{\omega^2({\bf k},
\lambda)/\varepsilon_s}{M_P/m^*} \\
\Omega^2_{int}({\bf k}, \lambda) &\simeq& \omega_{pol}^2+
\frac{\omega^2({\bf k}, \lambda)}{\varepsilon_\infty}
\end{eqnarray}
Those frequencies are sketched in figure \ref{fig-Eigen}. At low
density, there is a clear separation of energy scales between:

i) the low frequency ``external'' modes with $\Omega_{ext}\ll \omega_{LO}$,
which correspond to a Wigner crystal of particles with mass $M_P$
instead of $m^*$.  At such low frequencies, the polarons behave as rigid
particles and the internal electronic motion can be
neglected. Furthermore, all the polarization cloud participates in the
screening of the dipolar interactions,
so that the eigenfrequencies are reduced by a factor $\varepsilon_s$.

ii) the high frequency ``internal'' modes with $\Omega_{int}\gg \omega_{LO}$,
which correspond to a Lorentz lattice of dipoles with characteristic
frequency $\omega_{pol}$, which was extensively studied in
ref. \cite{cata}. In this case, the electron vibrations are very
fast and the appropriate dielectric constant is $\varepsilon_\infty$.

Such separation of energy scales is strictly valid only at very low
densities. At densities such that $\omega_{pol}$ approaches
$\omega_{LO}$, i.e. close to the phonon instability (see below), the full
expression (\ref{Eigen}) must be used to determine the eigenfrequencies.

The treatment of this section provides us with an estimate of the
low-lying excitations of the PWC. At this stage, it is useful to
comment on a few points concerning our approximations.
First of all, above some threshold  frequency higher than
$\omega_{pol}$, the calculated eigenmodes will eventually merge into a
continuum of states, corresponding to free electrons ionized from both
the polarization potential and the periodic potential of the Wigner
crystal itself. In addition, more bound states can appear below such
threshold, if there are metastable interstitial states with long
enough lifetimes.
As far as the Feynman approximation is concerned, we know from the single
polaron results \cite{devreese} that the interpretation of the
eigenfrequency of the quadratic model as representative of
the excitation spectrum of the real system
is qualitatively
correct in the strong coupling limit, where the optical  Franck-Condon
transition is very sharp (indeed, the polarizability of a system with
a definite quantum transition has exactly the same form as for
a classical harmonic oscillator).
At intermediate values of $\alpha$, however, the transitions inside
the polarization potential-well are considerably broadened. In this
case,
the calculated eigenfrequencies should still be of significance, provided
that we associate to them a finite linewidth $\Gamma$, related to the so-called
Huang-Rhys factor.

\subsection{Instability of the polaron system}
Contrary to the ordinary WC of electrons, the PWC can undergo a \textit{phonon
instability} when the density reaches a critical value.
This can be most clearly seen by writing
the action (\ref{Seff}) as a sum over the
quadratic actions of the independent eigenmodes ${u}_{\mathbf{k}\lambda}$,
$S_{eff}=\sum_{{\bf k}\lambda}S_{{\bf k}\lambda}$, with
\begin{eqnarray}
        S_{\mathbf{k}\lambda} &=& -\frac{m^*}{2}\int
        |\dot{u}_{\mathbf{k}\lambda}(\tau)|^2 d\tau
        - \frac{m^*\omega^2(\mathbf{k},\lambda) }
        {\varepsilon_s} \int |u_{\mathbf{k}\lambda}(\tau)|^2 d\tau
\label{S0-k}\\
        &-&\frac{K(\mathbf{k},\lambda)\omega_{LO}}{8}
        \int \! \int
        |u_{\mathbf{k}\lambda}(\tau)-u_{\mathbf{k}\lambda}(\sigma)|^2
        G_{\beta,\omega_{LO}} \, d\tau \, d\sigma
        \nonumber
\end{eqnarray}
We see that each of the $S_{{\bf k}\lambda}$ has the same form as the
mean-field  action (\ref{qaction}), but the coefficient of the polaron
term (the retarded interaction of ${u}_{\mathbf{k}\lambda}$ with
itself) is now
\[
K(\mathbf{k},\lambda)=K\left(1+\frac{1}{\tilde{\varepsilon}}\frac{m^*}{M}
        \frac{\omega^2(\mathbf{k},\lambda)-\omega_W^2}{\omega_{LO}^2} \right)
\]
Both $\omega^2 ( {\bf k},
\lambda)$ and $\omega_W^2$ are proportional to the density, so that
for $n\rightarrow 0$ we recover $K(\mathbf{k},\lambda)=K$.
However, upon increasing the density, the restoring force
$K(\mathbf{k},\lambda)$ can become negative, and the system is unstable.
The first instability occurs due to the
long wavelength transverse modes for which $\omega(k=0,t)=0$.
It is thus
\textit{the long
wavelength transverse vibrations of the electrons relative to their
polarization cloud which cause the instability}, a phenomenon which
is reminiscent of the phenomenological dissociation criterion introduced
in reference \cite{PWC}.

In terms of the parameter $\omega_{pol}$, we see that $K(k=0,t)$
vanishes
(i.e. the system is on the
verge of the instability) when
\begin{equation}
\omega_{pol}=\omega_{LO}
\label{inst}
\end{equation}
This instability condition is
the correct generalization of the condition $\omega_{pol}=0$ of
ref. \cite{cata} to the case of a finite phonon frequency.
Let us stress once again that the phonon instability occurs at $k=0$, where
the properties of the system are independent of the details of the
lattice ordering, provided that the system remains isotropic.
As a consequence, the instability criterion
(\ref{inst}) also applies to disordered insulating polaron phases,
which can be realized in real solids where  the doping ions act as
impurity potentials, thus driving the particles away from their
configurations on a Bravais lattice to form a Wigner glass. The above
criterion should also hold in the case of a  polaron liquid, provided that
the motion of the polarons is slow compared to the vibrations of the
electrons inside their potential wells \cite{note-liquid}.
\begin{table}[htb]
\centering{\begin{tabular}{|c|c|c|c|} \hline
$\alpha$ & $r_s^{(melt)}$ & $r_s^{(dis)}$ & $r_s^{(inst)}$  \\ \hline
0 & 960 & - & - \\
3 & 537 & - & 19 \\
7 &  52 & - & 18 \\ \hline \hline
10&  -   & 32& 16 \\

100& -   & 31& 15 \\
\hline
\end{tabular}}
\caption{The density parameter $r_s^{(inst)}$ for the phonon instability,
compared with the critical $r_s$ for crystal melting and polaron
dissociation in the mean field approximation (see ref. \cite{PWC}).
Parameters are $\varepsilon_\infty=5,\ \varepsilon_s=30$ and $m^*=2m_e$. The critical
densities can be obtained from the relation $n=1.6\cdot 10^{24}/r_s^3
cm^{-3}$.}
\end{table}

In Table I we have reported  the instability parameter $r_s^{(inst)}$
as deduced from criterion (\ref{inst}), compared to the values for
polaron dissociation and crystal melting obtained in reference
\cite{PWC}. In the weak coupling limit, the retarded interactions can
be neglected and the PWC tends to a WC of electrons, which has no
phonon instability \cite{bagchi}. Correspondingly, $r_s^{(inst)}$
vanishes (the critical density diverges) as $\alpha \rightarrow 0$. In
the opposite strong coupling limit, where $\omega_{LO}\rightarrow 0$, we
recover the results of ref. \cite{cata}. Generally speaking, in
all the range $\alpha>\alpha^*$, the results for $r_s^{(inst)}$  agree within a
factor of two with the critical $r_s$ for polaron dissociation, which
confirms that there is
an intimate connection between the instability of the
long wavelength transverse phonons and the mechanism of polaron
dissociation studied in previous references.
On the contrary, for $\alpha<\alpha^*$,
the crystal melts due to the quantum fluctuations of the polarons
 before the occurrence of the phonon instability, and the latter
is never realized in practice.

\begin{figure}
\centerline{\resizebox{7cm}{!}{\includegraphics{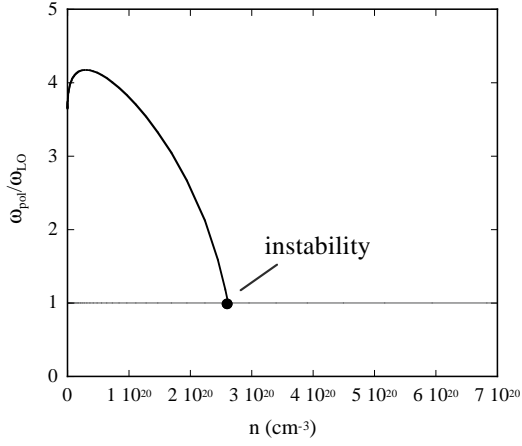}}}
\caption{Softening of the collective frequency $\omega_{pol}$ with
density. The instability occurs when
$\omega_{pol}=\omega_{LO}$. Parameters are $\alpha=7,\
\varepsilon_\infty=5,\ \varepsilon_s=30$ and $m^*=2m_e$.}
\label{peakshift}
\end{figure}

Figure \ref{peakshift} shows the evolution of $\omega_{pol}$ vs the density in
an intermediate situation, $\alpha=7$.
The behaviour is very similar to the case $\omega_{LO}=0$ presented in
ref. \cite{cata}, where it was shown that the frequency of the
collective mode is gradually reduced due to the effects of the local
Lorentz field (see also ref. \cite{lorenzana,note-upturn}).

\section{Longitudinal Dielectric Constant and Optical Response.}
\label{dielectric constant}
In this section, we derive the screening properties and the absorption
spectrum of the PWC in the harmonic approximation.
The determination of the dielectric constant of the model (\ref{Leff})
follows the lines given in ref. (\cite{bagchi}), and the optical
conductivity is deduced from the $k=0$ limit of this quantity.

\subsection{ Dielectric Constant}
The dielectric constant is defined as the response to a perturbing
external charge, that we take  of the form
\begin{equation}
\rho_{ext}({\bf r},t)=\rho_0 e^{i({\bf k\cdot r}-\omega t)}
\end{equation}
Taking into account the screening of the host polar lattice, which we
schematize as $\varepsilon_{ph}(\omega)=
\varepsilon_\infty (\omega^2-\omega_{LO}^2)/(\omega^2-\omega_{TO}^2)$,
$\omega_{LO}$ and $\omega_{TO}$ being respectively the longitudinal
and transverse optical phonon frequency, the force acting on an
electron at site $i$ can be written as
\begin{equation}
{\bf F}_i(t)=-e{\bf E}_{ext}({\bf{R}_i,t})=i\frac{4\pi {\bf k}}{k^2}
\frac{e\rho_0}{\varepsilon_{ph}(\omega)}e^{i({\bf k\cdot r}-\omega t)}
\end{equation}
Introducing this driving force into the equations of motion
(\ref{eqmot}) and solving for the induced displacement $u_{{\bf
k}\lambda}$ we obtain
\[
u_{{\bf k}\lambda}=  \frac{i4\pi e\rho_0}{\varepsilon_{ph}(\omega)}
\frac{({\bf k\cdot \epsilon_{{\bf k}\lambda}})^2}{k^2}
\frac{\omega_{LO}^2-\omega^2}
{\left[ \Omega^2_{int}({\bf k}, \lambda) -\omega^2 \right]
\left[ \Omega^2_{ext}({\bf k}, \lambda) - \omega^2 \right]}
\]
The longitudinal dielectric constant can be obtained by comparing the
total charge associated to the displacement $u_{{\bf k}\lambda}$ and the
perturbing charge distribution, namely
$\varepsilon_L({\bf k}, \omega)=\rho_{ext}/\rho_{tot}$
(for more details, see ref. \cite{bagchi,cata}).
The result for the PWC reads
\begin{eqnarray}
\frac{1}{\varepsilon_L({\bf k},
\omega)}&=&\frac{1}{\varepsilon_{ph}(\omega)}
\left\lbrace
1+\frac{\omega_p^2}{\varepsilon_\infty}\sum_\lambda
\frac{({\bf k\cdot \epsilon_{{\bf k}\lambda}})^2}{k^2} \times \right.
\label{diel}
 \\
& & \times \left. \frac{\omega^2-\omega_{TO}^2}
{\left[ \Omega^2_{int}({\bf k}, \lambda) -\omega^2 \right]
\left[ \Omega^2_{ext}({\bf k}, \lambda) - \omega^2 \right]}
\right\rbrace
\nonumber
\end{eqnarray}
This expression can be simplified in the following limits:
\subsubsection{long wavelengths ($k=0$)}
\begin{figure}[ht]
\resizebox{8cm}{!}{\includegraphics{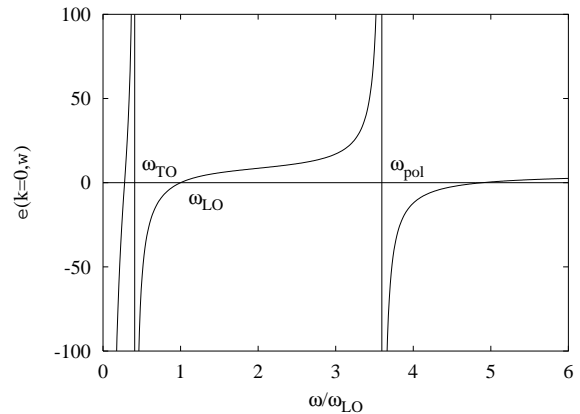}}
\caption{Dielectric function of the PWC at $k=0$, with the same
parameter set as in figure \ref{peakshift} at
$n=5 \cdot 10^{19} cm^{-3}$. From the
Lyddane-Sachs-Teller relation, the transverse phonon frequency is
$\omega_{TO}=0.41 \omega_{LO}$.}
\label{fig-diel-k0}
\end{figure}

Since in this limit  ${\bf k}\cdot \epsilon_{{\bf k} t}=0$,
the transverse modes do not enter in the sum over
$\lambda$, and the dielectric constant reduces to
\begin{equation}
\varepsilon_L(0,
\omega)=\varepsilon_{ph}(\omega)
\frac
{\left[ \omega^2 -\Omega^2_{int}(0, \ell)  \right]
 \left[ \omega^2 -\Omega^2_{ext}(0, \ell)  \right]}
{\omega^2 \left[\omega^2-\omega_{pol}^2\right]}
\label{diel-k0}
\end{equation}
This result is sketched in figure \ref{fig-diel-k0}.
In this limit, the transverse and longitudinal dielectric
constants become equal, and the zeros and poles of (\ref{diel-k0})
correspond respectively to the longitudinal and transverse collective modes.

The pole at $\omega=0$ corresponds to a transverse sliding mode
of the PWC, which is equivalent to the sliding mode of a CDW
\cite{LRA}. This mode is
expected to be current-carrying when the PWC is not pinned, as is the
case in our jellium model, leading to an infinite d.c. conductivity.
In real materials, however, this excitation
can be shifted to higher frequencies due to the pinning by
impurities and commensurability with the host lattice. As a
consequence, the infinite d.c. conductivity is desrtoyed and the
infinite (negative) dielectric constant at $\omega=0$ is replaced by a
large (positive) value below some threshold frequency $\Delta_{pin}$
characteristic of the pinning process (this effect can be introduced in our
model in a phenomenological way by taking the frequency
$\Omega_{ext}({\bf k},\omega)$ of the transverse mode to be equal to
$\Delta_{pin}$ at low $k$).

\subsubsection{static limit ($\omega=0$)}
In this limit, eq. (\ref{diel-k0}) becomes
\begin{equation}
\frac{1}{\varepsilon_L({\bf k},0)}=\frac{1}{\varepsilon_s}
\left\lbrace
1-\frac{\omega_p^2}{k^2}\sum_\lambda
\frac{({\bf k\cdot \epsilon_{{\bf k}\lambda}})^2}
{\omega^2({\bf k}, \lambda)}
\right\rbrace
\label{diel-w0}
\end{equation}
By comparing this expression with the result of ref. \cite{bagchi},
we see that the static response of the PWC in the jellium model
is identical to the
ordinary WC, with an enhancement $\varepsilon_s$ due to the host
phonon screening.

\subsubsection{sign of the dielectric constant}

By taking a simple model for the phonon dispersion of the WC of
electrons, we can study the sign of the longitudinal dielectric
constant of the PWC in the $({\bf k},\omega)$ space. The model
consists of two transverse acoustical branches $\omega({\bf
k},t)=ak$ and one longitudinal optical branch $\omega({\bf
k},l)=\sqrt{\omega_p^2-2a^2k^2}$, in order to fulfill the Kohn sum
rule. An estimate of the parameter $a$ (the velocity of sound)
can be obtained by arguing that the value of the transverse
frequency at the end of the Brillouin zone
($k_0=\pi/R_s$), must be equal to some fraction
of the plasma frequency (we shall take $a=0.1  \omega_p/k_0$).
Substituting these expressions into eq. (\ref{Eigen}) we get the
eigenfrequencies of the PWC, that we use in eq. (\ref{diel}) to determine the
dielectric constant as a function of the density.
For simplicity, we restrict ${\bf k}$ to vary along one of the
principal axes of the crystal, where by symmetry
the transverse polarisation
vectors obey $\epsilon_{{\bf k}t}\cdot {\bf k}=0$.
\begin{figure}
\resizebox{9cm}{!}{\includegraphics{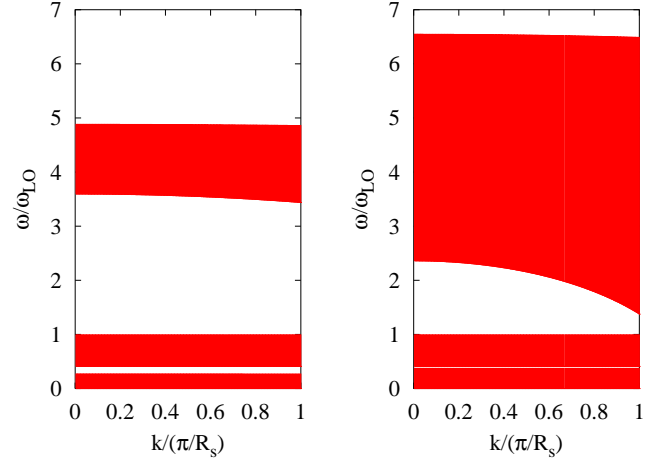}}
\caption{Regions in the $(k,\omega)$ plane where the dielectric
constant is negative, which are located between the zeros and the
poles of $\varepsilon_L$. Left panel: $n=5 \cdot 10^{19} cm^{-3}$;
right panel: $n=1.7 \cdot 10^{20} cm^{-3}$. Parameters are the same as
in figure 1.}
\label{fig-diel-sign}
\end{figure}

As can be seen in figure \ref{fig-diel-sign}, the regions where
$\varepsilon_L<0$ expand with increasing density. When one approaches
the critical density, the dielectric constant is negative in all the
region below the frequency $\Omega_{int}(k,\ell)\simeq
\omega_p/\sqrt{\varepsilon_\infty}$.

\subsection{ Optical Conductivity}
\begin{figure}[t]
\resizebox{8cm}{!}{\includegraphics{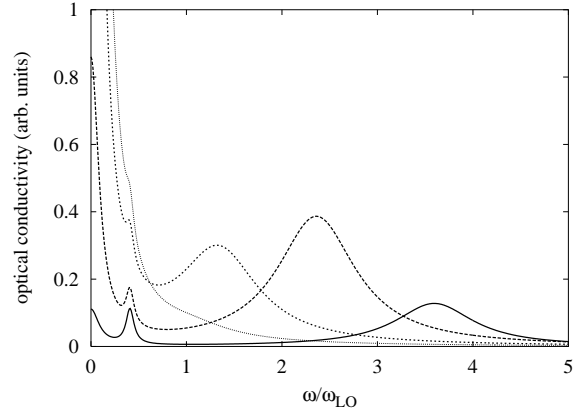}}
\caption{Optical conductivity of the PWC at various densities:
$n=0.5, \ 1.7, \ 2.3$ and $2.4\cdot 10^{20} cm^{-3}$, with the same
parameter set as in figure \ref{peakshift} (the instability occurs at
$n^{(inst)}=2.7 \cdot 10^{20} cm^{-3} $). We have introduced the
phenomenological linewidths $\Gamma_{pol}= \omega_{LO}$,
$\Gamma_{0}=\Gamma_{ph}=0.1 \omega_{LO}$. Note the shift of the
polaron peak, and  the transfer of spectral weight taking place
between the feature at $\omega_{pol}$ and the zero frequency peak. 
When the polaron lattice is pinned by impurities or by
commensurability effects, the $\omega=0$
sliding mode moves to a finite frequency $\omega=\Delta_{pin}$.}
\end{figure}
The real part of the
optical conductivity is related to the imaginary part of the
transverse dielectric constant by
\begin{equation}
\sigma(\omega) = \frac{\omega}{4\pi} Im \  \varepsilon(k=0,\omega)
\end{equation}
We remark that expression (\ref{diel-k0}) is perfectly real, and
therefore it can only lead to delta-like absorption peaks located at
its poles  $\omega=0$ and $\omega=\omega_{pol}$.
As was stated in the previous section, at finite values of
the e-ph coupling, the collective excitations have an intrinsic
lifetime which cannot be obtained within the quadratic model
(\ref{Leff}).  We thus introduce phenomenological linewidths by
putting $\omega\rightarrow \omega+i\Gamma$. Substituting the values of
$\Omega_{int}(0,\lambda)$ and $\Omega_{ext}(0,\lambda)$ given by
eq. (\ref{Eigen}), we obtain
\begin{equation}
\sigma(\omega) =\sigma_{ph}(\omega)+\sigma_{pol}(\omega)+ \sigma_0(\omega)
\end{equation}
where  the first term is the response from the phonons of the host
medium, the second term comes from the collective transverse
excitation at $\omega_{pol}$, and the last term is the contribution
from the sliding mode centered at
$\omega=0$. As was stated in the previous section, 
if the PWC is pinned due to disorder or lattice commensurability, 
a gap opens in the optical conductivity, and the zero-frequency peak 
is shifted to $\omega=\Delta_{pin}$. Therefore,  the optical response
of the polaron Wigner crystal is characterized by  the emergence of
two collective-mode peaks, a polaron peak located above the host
phonon frequency
and a sliding (or pinning) peak around or  below the phonon frequency.

Assuming that the above
contributions are well
separated  from one another (i.e. that the peaks do not overlap),
these take the form of Lorentz peaks:
\begin{eqnarray}
 \sigma_{ph}(\omega)&=&\frac{\varepsilon_s-\varepsilon_\infty}{4\pi}
 \omega_{TO}^2
 \frac{\Gamma_{ph}/4}{(\omega-\omega_{TO})^2+\Gamma_{ph}^2/4} \\
\sigma_{pol}(\omega)&=& \left( 1- \frac{\omega_{LO}^2}{\omega_{pol}^2}\right)
\frac{\omega_p^2}{4\pi}
 \frac{\Gamma/4}{(\omega-\omega_{pol})^2+\Gamma^2/4}
\label{sig-pol}\\
\sigma_0(\omega)&=&\left( \frac{\omega_{LO}^2}{\omega_{pol}^2}\right)
\frac{\omega_p^2}{4\pi} \frac{\Gamma_0}{\omega^2+\Gamma_0^2} \label{sig-0}
\end{eqnarray}
The total spectral weight associated with the absorption of the PWC is
$\omega_p^2/8$, and it obeys the conductivity
sum rule. At low
density, if $\alpha>\alpha^*$, one typically has $\omega_{pol}\gg
\omega_{LO}$ and almost  all the spectral weight is carried by the
polaron peak in $\sigma_{pol}(\omega)$, as can be seen from eq.
(\ref{sig-pol}). As
the density is increased, some spectral weight is transferred to the
sliding mode at $\omega=0$ ($\omega=\Delta_{pin}$ in the pinned
case), until the polaron peak eventually disappears at
the instability point.

\section{Conclusion}
\label{conclusion}

In this paper, we have analyzed the melting of a polaron Wigner
crystal in the strong electron-phonon coupling limit, taking into
account the dipole-dipole interactions between localized electrons,
i.e. going beyond the mean field appropach developed in
ref. \cite{PWC}.  We have shown that the system undergoes a phonon
instability due to the effect of the Lorentz local field, which
confirms the polaron dissociation scenario introduced previously.
The instability has an optical signature: the polaron peak
in the optical conductivity \textit{shifts towards low frequencies} 
as the doping density is increased. 
This red-shift should be accompanied by the rise of a 
collective peak in the far infrared, whenever the polaron Wigner crystal is
pinned by disorder or by commensurability effects.
This complex behaviour has been
experimentally observed in NCCO \cite{calvani}, which suggests
that \textit{both} the polarons and the unscreened long-range 
Coulomb interactions should play an important role in the high Tc
compounds. We suggest that the same behaviour could be observed
 in other superconducting materials.

We have calculated the dielectric constant of the PWC, and
shown that it is negative on a large region of $({\bf k},
\omega)$. This indicates the possibility of reaching a 
superconducting phase, if carriers of different nature 
--- localized polarons and free electrons --- can coexist close to the 
instability density. In such a regime, the itinerant
electrons would be paired through the high frequency internal vibrations
of the polaron lattice, which would act as the phonons in the
ordinary BCS theory. The large collective frequencies of the polaron
crystal would naturally lead to a high superconducting transition temperature.
Eventually, at higher doping levels, the polaron
clouds are completely screened by the free electrons, so that the mixed
phase (and the superconducting phase) should disappear in favor of a 
more conventional metal.\cite{note-alex}

It should be kept in mind that
our results are based on the continuous Fr\"ohlich model in the low
density regime and at strong electron-phonon coupling, which 
misses many essential physical features of the cuprates.
Indeed, we are not considering the spin degrees of freedom,
although these are certainly responsible for most of the anomalous
properties of the normal state.
We are not taking into account the formation 
of stripes, although the practical realization of the mixed phase
conjectured above could 
well involve such form of microscopic phase segregation.
On the other hand, the inclusion of 
other ingredients, such as the anisotropy of
electronic bands in the cuprates, or the discretization of the lattice, 
would increase the tendency to localization and are therefore
 expected to push the system towards the 
present ''polarization catastrophe'' scenario.

The authors would like to thank P. Calvani, B.K. Cha\-kra\-verty,
S. Ciuchi, D. Feinberg, 
D. Mayou and J. Ranninger for many stimulating discussions.


\begin{thebibliography}{}
\bibitem{note-liquid} In that case, one
should use in formula (\ref{wpol}) the values of the parameter
$M$ appropriate for a polaron liquid.
\bibitem{note-upturn}
Concerning the upturn observed at very low
density, a possible explanation is that, before the dipolar
interactions become effective in reducing the collective frequency
$\omega_{pol}$,  the polarons tend to shrink
their radius owing  to the repulsion of their nearest neighbors, so
that the  characteristic vibration frequency increases (indeed, the
mean field value of $M$ that we have used in eq. (\ref{wpol})
increases with the density, as can be seen in fig. 5 of ref. \cite{PWC}).
However, this point is quite subtle and should be clarified by solving
the problem in a fully
self-consistent way, as stated in section \ref{approximations}.
\end{thebibliography}
\end{document}